
\input harvmac.tex

\def\frac#1#2{{#1\over#2}}

\def\exp{{\rm exp}}

\def\slash#1{\mathord{\mathpalette\c@ncel{#1}}}
\overfullrule=0pt

\def\steepslash{\c@ncel}
\def\frac#1#2{{#1\over #2}}

\def\C{{\bf C}}
\def\inbar{\,\vrule height1.5ex width.4pt depth0pt}
\def\IB{\relax{\rm I\kern-.18em B}}
\def\IC{\relax\hbox{$\inbar\kern-.3em{\rm C}$}}
\def\IP{\relax{\rm I\kern-.18em P}}
\def\IR{\relax{\rm I\kern-.18em R}}
\def\IZ{\relax\ifmmode\mathchoice
{\hbox{Z\kern-.4em Z}}{\hbox{Z\kern-.4em Z}}
{\lower.9pt\hbox{Z\kern-.4em Z}}
{\lower1.2pt\hbox{Z\kern-.4em Z}}\else{Z\kern-.4em Z}\fi}

\catcode`\@=12

\Title{\vbox{\baselineskip12pt\hbox{RU-93-55}
                \hbox{hepth@xxx/9311189}}}
{\vbox{\centerline{
Correlators Of The Jost Functions}
\vskip6pt\centerline{ In The Sine-Gordon Model}}}

\centerline{Sergei Lukyanov \footnote{$\dagger$}
{On leave of absence from L.D. Landau Institute for Theoretical
Physics, Kosygina 2, Moscow, Russia}
\footnote{$^*$}{e-mail address: sergei@physics.rutgers.edu}}
\centerline{Department of Physics and Astronomy}
\centerline{Rutgers University, Piscataway, NJ 08855-049}
\centerline{}
\centerline{}

\centerline{\bf{Abstract}}
In this paper the quantum direct scattering problem is solved for
the Sine-Gordon model.
Correlators of the  Jost functions are derived by the
angular quantization  method.
\Date{November, 93}

\newsec{Introduction}
The Inverse Scattering  Method
\ref\ggkm{C.S. Gardner, T.M. Green, M.D. Kruskal and
R.M. Miura, Phys. Rev. Lett. 19 (1967) 1095},
\ref\col{M. Ablowitz,
P. Kaup, A. Newell and H. Segur, Phys. Rev. Lett. 30
(1973) 1262},
\ref\fa{L.D. Faddeev and L.A. Takhtajan, Hamiltonian Method in the
Theory of Solitons, Springer, N.Y. (1987)}
is the most powerful
technique to solve non linear differential equations.
It consists of two essential steps. First,  we have
to  transform  initial conditions to the  scattering
data which are  the Jost functions of the auxiliary linear
problem. One can consider these functions as
an analogue of  proper regularized Wilson P-ordered exponents
in gauge theories. In terms of Jost functions the
dynamics  becomes trivial. Hence, the solution of the initial
equation is reduced to  inverting  of the scattering transform.

In  the quantum case  handling an equation  means
a reconstruction of all correlators of local fields.
{}From this point of view the solution of the quantum direct
scattering problem
consists of reconstructing  all correlators
of the Jost functions. In the present paper  this problem
is solved
for the Sine-Gordon model.

The  approach we are dealing with in this paper
\ref\zam{A.B. Zamolodchikov, unpublished}
can be considered as a proper
generalization of the Baxter transfer matrix  method
for statistical systems
\ref\bax{R.J. Baxter,
Exactly Solved Models in Statistical Mechanics,
Academic Press, London (1982) },\
\ref\japtwo{B. Davies, O. Foda, M. Jimbo, T. Miwa
and A.Nakayashiki, Comm. Math. Phys. 151 (1993) 67;\
M. Jimbo, K. Miki, T. Miwa and A. Nakayashiki,
Phys. Lett. A168 (1992) 256.}.
Notice that an analogue of the Jost functions for lattice systems is
the first kind of vertex operators introduced in the papers\ \japtwo.
Correlators of the Jost functions can be expressed
in terms of  traces over the space of angular quantization.
The latter is the scaling limit of the space
where the lattice corner transfer matrix acts. The angular
quantization space for the Sine-Gordon model
has been constructed in\
\ref\shat{S. Lukyanov and S. Shatashvili, Phys. Lett. B298 (1993) 111},
\ref\luk{S. Lukyanov, Free Field Representation for Massive
Integrable Models,
Rutgers preprint RU-93-30 (1993) (hep-th/9307196)}.
Here it is  shown that the transcendental  functions
introduced in \ \luk\   are nothing but  correlators of the
quantum Jost functions.

The paper is  organized as follows.
We explain the main ideas
of the construction  by the elementary example of
the Klein-Gordon equation\
\ref\hgf{S.A. Fulling, Phys. Rev. D7,\ 4 (1973) 2850},
\ref\hdgf{W.G. Unruh, Phys. Rev. D14, 4 (1976) 870; W.G. Unruh
and R.M. Wald, Phys. Rev. D29, 6 (1984) 1047}.
Then they are applied to the Sine-Gordon model.

\newsec{The Klein-Gordon model}

Let us start with  a consideration of the Klein-Gordon equation
\eqn\sine{(\partial_t^2-\partial_x^2)\varphi+m^2\varphi=0\ .}
The phase space of the model is generated by the canonical
variables
\eqn\hdy{\eqalign{\pi(x)=\partial_t&\varphi(x,t)|_{t=0},
\ \ \ \varphi(x)=
\varphi(x,t)|_{t=0},\cr
&\{\pi(x),\varphi(y)\}=\delta(x-y)\  .}}
We shall suppose that the functions\ $\pi(x),\varphi(x)$\
decrease at infinity fast enough
and admit decomposition in Fourier integrals:
\eqn\hsghs{\eqalign{
&\varphi(x)=2^{-\frac{1}{2}}\int _{-\infty}^{+\infty}\frac{d \beta}
{2\pi}\big(A^{*}(\beta) e^{-i m x \sinh\beta}+
A(\beta) e^{i m x \sinh\beta}\big ) ,\cr
&\pi(x)=2^{-\frac{1}{2}}i\  m \int _{-\infty}^{+\infty}\frac{d \beta}
{2\pi}\ \cosh\beta\
\big (A^{*}(\beta) e^{-i m x \sinh\beta}-
A(\beta) e^{i m x \sinh\beta}\big )\ .}}
The solution of \ \sine
\  satisfying the initial data\ \hsghs\
has the form
\eqn\jku{\varphi(x,t)=2^{-\frac{1}{2}}
\int _{-\infty}^{+\infty}\frac{d \beta}
{2\pi}\big (A^{*}(\beta) e^{-i m ( x \sinh\beta-t \cosh \beta)}+
A(\beta) e^{i m ( x \sinh\beta-t \cosh \beta)}\big )\ .}
Let us fix the point \ $ (x_0,t_0)=0$\ and introduce
the Rindler coordinates
\eqn\jshg{\eqalign{
x=r \cosh \theta,\ \  & \ \ t= r \sinh \theta;\cr
-\infty<\theta<+\infty,\ \ &\ \ \ 0<r<+\infty\  ,}}
in the
space-like region
\eqn\gfbn{x>|t|>0\ .}
For any ray \ $\theta=const$\
we define the map \ $$Scat:\  \varphi(r,\theta)
\rightarrow \lambda^{\theta}(\alpha)$$
in the following way:
\eqn\jhggg{\lambda^{\theta}(\alpha)=\lim_{a\rightarrow 0}
\int _{a}^{+\infty} d r
e^{i m r \sinh \alpha}
\big (-\frac{1}{r}\partial_{\theta}+i m \cosh\alpha
\big ) \varphi(r,\theta)\ .}
Using the formula\ \jku\  it is easy to show  that  the dynamics
becomes trivial in the variables
\ $\lambda(\alpha)\equiv \lambda^{\theta=0}(\alpha)$:
\eqn\kdi{\lambda^{\theta}(\alpha)=\lambda(\alpha-\theta)\ .}
This observation allows one to solve the Klein-Gordon equation
by applying the direct and inverse transformation  \jhggg\ .
The  transformation is nothing but  Laplace one and
it is invertible in the class of decreasing functions.
So, the solution
\ $\varphi (x,t)$\ in the space-time region  \gfbn\
is uniquely defined by the initial
data\ $\pi(x), \varphi(x)$ \ for $x>0$.
It might be useful to point out
that this solution method of the Klein-Gordon
equation is a prototype of the inverse scattering  technique
for integrable equations.

Let us recall now essential steps of the
quantization of the Klein-Gordon equation.
The canonical Poisson structure\ \hdy\  can be quantized by
using the correspondence principle, which
prescribes replacing the classical Poisson
bracket with the commutator\ $\frac{i}{\hbar}[\ ,\ ]$.
All required commutation relations
follow  immediately from the canonical commutators. In particular,
\eqn\ksjs{[\lambda(\alpha_1),\lambda(\alpha_2)]=i\ \hbar
\ \tanh\frac{\alpha_1-
\alpha_2}{2}\ .}
After the quantization
the complex conjugation * becomes the Hermitian  one\ $ {}^+$.
The Hilbert space of the theory
is generated by the basic vectors
\eqn\lksj{\eqalign{|A(\beta_1)...A(\beta_n)>_{in}
&=A(\beta_n)...A(\beta_1)|vac>,\cr
\beta_n>\beta_{n-1}&>...>\beta_1\ .}}
Here the vacuum state
$|vac>$ is specified by the conditions:
\eqn\vac{A(\beta)|vac>=0,\ \ \ <vac|A^{+}(\beta)=0\ . }
In what follows we shall need the explicit form of
the vacuum average
\eqn\gsfd{F(\alpha_1-\alpha_2)=
<vac|\lambda(\alpha_2)\lambda(\alpha_1)|vac> .}
Using the formulas\ \jku,\jhggg,\vac\  one can get the relation:
\eqn\ksjjs{F(\alpha_1-\alpha_2)=\frac{\hbar}{4}
\int _{-\infty}^{+\infty}\frac{d \beta}
{2\pi}\coth\frac{\beta+\alpha_1+i 0}{2}
\tanh\frac{\beta+\alpha_2}{2}\ .}
The integral \ \ksjjs\ is divergent. To assign a meaning to it
we introduce the ultraviolet cut-off
\eqn\jhhg{-\frac{\pi}{\epsilon}
<\beta<\frac{\pi}{\epsilon},\ \  \ \ \epsilon
\rightarrow 0\  .}
Then
\eqn\jshsgf{F(\alpha)=
-\frac{\hbar}{2\pi}(\alpha+i\pi)\tanh\frac{\alpha}{2}+
\frac{\hbar}{4 \epsilon}+o(\epsilon)\ .}
The reason of this divergence is simple. It indicates that
the limit \ $a\rightarrow 0$\ in the quantum definition\ \jhggg\
demands  regularization. The parameters\  $a$\ and
\ $\epsilon$ are related by the obvious relation
\eqn\kkjju{\frac{\pi}{\epsilon}\sim \ln m a\   .}

The explicit form of the function \ $F(\alpha)$ \
makes it possible to reconstruct correlation functions
of any operators. For instance, as  follows from
definition\ \jhggg\ ,
\eqn\jshgd{\eqalign{&\int _{0}^{+\infty}
\int _{0}^{+\infty}d r_1 d r_2
e^{i m (r_1\sinh\alpha_1+r_2\sinh\alpha_2)}
<vac|\varphi(r_1,\theta_1)
\varphi(r_2,\theta_2)|vac>=\cr&
= (m^2\pi\cosh\alpha_1\cosh\alpha_2)^{-1}
\big[F(\theta_1-
\theta_2+\alpha_1+\alpha_2-i\pi)+F(\theta_1-
\theta_2-\alpha_1-\alpha_2+i\pi)-\cr
&\ \ \ \ \ \ \ \ \ \ \ -F(\theta_1-
\theta_2-\alpha_1+\alpha_2)-F(\theta_1-
\theta_2+\alpha_1-\alpha_2)\big],\ \ \ \theta_1>\theta_2\  .}}
Inverting the Laplace transformations
\ \jshgd\  one can get the familiar expression for
the $T$-ordered Green's function \ $<vac|T[\varphi(x_1,t_1)
\varphi(x_2,t_2)]|vac>.$
Of course, it does not depend on the regularization
parameter\ $\epsilon$\ \jhhg.

Since the operator\ $\lambda(\alpha)$\ depends only
on\ $\pi(x),\ \varphi(x),\ x>0$,  it would seem that
we reconstructed  the Green's function only
in terms of the Hilbert space\ $\pi_Z$ associated with the
half infinite line\ $x>0, t=0$\ .
But this is not correct.
The point is that  the vacuum vector\ $|vac>$\ \vac\  does not
belong to the space\ $\pi_Z$\ .
Contrary to the classical situation, the quantum dynamics
in the area\ \gfbn\  depends on  quantum fluctuations penetrating
through the light cone. Hence, the function\ \gsfd\
can be expressed
only in terms of  some density matrix\ $\hat\rho$\
which controls these effects
\ \hdgf
\eqn\fsda{<vac|\lambda(\alpha_2)\lambda(\alpha_1)|vac>=
tr_{\pi_Z}\big[\hat \rho\
\lambda(\alpha_2)\lambda(\alpha_1)\big]\ .}
The reconstruction of the density matrix
is based on  the following
 property of the function\ $F(\alpha)$\ \jshsgf
\eqn\lsj{F(\alpha)=F(-2\pi i-\alpha)\ .}
Consequently,
\eqn\ksjhhd{\hat \rho=e^{2\pi i K} \ .}
Here $K$ is the operator of  Lorentz rotations
\eqn\hgfds{\eqalign{&e^{\theta K}\lambda(\alpha)e^{-\theta K}
=\lambda(\alpha-\theta)\ ,\cr
&K=
\frac{i}{2\hbar}\
\int_0^{+\infty}d r\ r\
[\pi (r)^2+\partial_r\varphi (r)^2+m^2\varphi (r)^2]\  .}}

It is necessary to point out that the
formula
\eqn\hsyfd{<vac|O_1(x_1,t_1)...O_n(x_n,t_n)|vac>=
tr_{\pi_Z}\big[e^{2\pi i K}O_1(x_1,t_1)...O_n(x_n,t_n)\big],}
where \ $O_k$\ are arbitrary operators,
is a universal relation. It is the condition of
uniqueness of $T$-ordered Euclidean Green's functions.
For a nontrivial quantum model a
description of the physical vacuum
is a difficult problem
\ref\far{L.D. Faddeev,
E.K. Sklyanin and L.A. Takhtajan, Theor. Mat. Fiz. 40
(1979) 194;\ Theor. Math. Phys. 40 (1979) 688}.
So we shall
consider the formula \ \hsyfd\
as a definition of the physical vacuum
state\ $|vac>$. It will be a crucial point for our construction.
Note that an analogous
idea was used to  investigate the
Hawking radiation in the Kruskal coordinates\ \hdgf .

Let us now discuss the structure of the space\ $\pi_Z$,
which will be called the space of angular quantization
\ \hgf ,
\hdgf .
The general solution of the Klein-Gordon equation in
the region  \ \gfbn\  has the  form:
\eqn\lopa{\varphi(r,\theta)= \int_0^{+\infty}\frac{d \kappa}{\pi}\
{\rm K}_{i\kappa}(mr)
[b_{\kappa} e^{-i \kappa \theta}
+b_{\kappa}^{+} e^{i \kappa\theta}]\ .}
Here \ ${\rm K}_{i\kappa}(mr)$\
are the Macdonald functions (modified
Bessel function) of imaginary order.
The function\ $\lambda(\alpha)$\  \jhggg\
and the operator of Lorentz
rotations\ $K$\  \hgfds\  can be expressed in
terms of the oscillators\ $b_{\kappa},
\ b_{-\kappa}\equiv b^{+}_{\kappa}\
(\kappa>0)$
\eqn\jdhf{\eqalign{&\lambda(\alpha+i\frac{\pi}{2})
= i\ \int_{-\infty}^{+\infty}d\kappa\
\frac{b_{\kappa}}{\sinh\pi \kappa}\ e^{i\kappa\alpha},\cr
&K=\frac{i}{\hbar}\
\int_{0}^{+\infty}d\kappa\ \frac{\kappa}{\sinh\pi\kappa}
b_{-\kappa}b_{\kappa}\ .}}
As it follows from the  formulas\ \ksjs,\jdhf ,
the oscillators\ $b_{\kappa}$\
satisfy the commutation relations
\eqn\hsfgd{[b_{\kappa},b_{\kappa'}]
=\hbar \sinh\pi\kappa\ \delta(\kappa+
\kappa')\ .}
The space of angular quantization\
$\pi_Z$\  is spanned by the  vectors
\eqn\lsjs{b_{\kappa_1}...b_{\kappa_n}|0>\in\pi_Z\ ,
\ \ \kappa_j<0\  .}
Here the vacuum vector\ $|0>\in\pi_Z$
(not to be confused with the physical vacuum\ $ |vac>$)
obeys the equations:
\eqn\jhdg{b_{\kappa}|0>=<0|b_{-\kappa}=0,\ \kappa>0\ .}
Using the oscillator representation of the space\ $\pi_Z$\
one can check the formula \ \fsda\    directly .
The
calculations have been done for the more general case of
the Sine-Gordon model
in the work\ \luk .

At the end of this section let us discuss the property
of the two point average
\eqn\jshd{f(\alpha_1-\alpha_2)=
<0|\lambda(\alpha_2)\lambda(\alpha_1)|0>\ .}
One can get an explicit form for this function:
\eqn\hggd{f(\alpha)=
\hbar\ \pi\ \psi(\frac{1}{2}+\frac{i\alpha}{2\pi})
+const\  .}
Here
$\psi(z)$ is the logarithmic derivative of the gamma function.
The function\ $f(\alpha)$\  satisfies the
conditions:

a. Due to the commutation relation\ \ksjs , it
obeys the functional equation
\eqn\uhhg{f(\alpha)-f(-\alpha)=
i \ \hbar \tanh\frac{\alpha}{2}\ .}

b. Due to the vacuum definitions\ \jhdg ,
\ $f(\alpha)$\   is an analytical function in the half plane
\ $\Im m\ \alpha\leq 0$\ .

c. If\ $Re\  \alpha\to\pm\infty\ $,
then\
$\partial_{\alpha}  f(\alpha)\to 0\ .$

\noindent
The last requirement has a  simple physical meaning. Indeed,
from \ \jhggg,\kdi\ it  follows  that it is equivalent
to the  boundary conditions
\eqn\ldjf{\eqalign{
&\partial_\theta\varphi(r,\theta)|0>
\to 0,\ \ \theta\to-\infty,\cr
&<0|\partial_\theta\varphi(r,\theta)\to 0,
\ \ \theta\to+\infty ,}}
which can be considered as conditions of  absence of the flow
through the light cone.

It is easy to see that the
requirements (a-c)  uniquely determine
the explicit form of
\ $f(\alpha)$\ \hggd\  and  the  whole  structure
of the space \ $\pi_Z$\ .
They admit  natural generalization
for the Sine-Gordon model and  can be used as the basis for
a reconstruction of the space of angular quantization\ \luk\ .

\newsec{The Sine-Gordon model}

The simplest integrable  generalization of
the  Klein-Gordon equation
is the Sine-Gordon model
\eqn\sie{(\partial_t^2-\partial_x^2)
\varphi+\frac{m^2}{b}\sin b\varphi=0,}
where \ $b$\ is the interaction constant.
The equation \ \sie\  can be represented as the zero-curvature
condition\ \col ,\fa\ :
\eqn\lks{[\partial_{\mu}-{\cal A}_{\mu},
\partial_{\nu}-{\cal A}_{\nu}]=0\   .}
It provides the classical integrability of \ \sie .
We shall consider
the solution of the Sine-Gordon equation in the
region\ \gfbn . Then,  in the
Rindler coordinates\ \jshg\  the connection
\ ${\cal A}_{\mu}$\ has the form:
\eqn\kajhgf{\eqalign{&{\cal A}_r
=\frac{1}{4 i}\big[ \frac{b}{r} \partial_{\alpha}
\varphi \sigma_3+2 m \cosh(\theta+\alpha)
\sin \frac{b \varphi}{2} \sigma_1+
2m \sinh(\theta+\alpha)
\cos \frac{b \varphi}{2} \sigma_2\big],\cr
&{\cal A}_{\theta}=\frac{r}{4 i}\big[ b \partial_{r}
\varphi \sigma_3+2 m \sinh(\theta+\alpha)
\sin \frac{b \varphi}{2} \sigma_1+
2m \cosh(\theta+\alpha)
\cos \frac{b \varphi}{2} \sigma_2\big].}}
Here \ $\sigma_k \ (k=1,2,3)$ are
Pauli matrices and \ $\alpha$ is a spectral
parameter.

The phase space of the model is generated  by
the canonical variables\ \hdy .
We shall discuss the decreasing boundary conditions
\eqn\hdg{\eqalign{&
\pi(r)=\frac{1}{r}\partial_{\theta}
\varphi (r,\theta)|_{\theta=0}
\to 0\ ,\cr
&\varphi(r)=\varphi (r,\theta)|_{\theta=0}
\to \frac{\pi l}{b}\ ,\ \ \ \ \ \  r\to+\infty\ .}}
Contrary to the free massive model,
the phase space contains
an infinite number of components
specified by the integer \ $l$\ .
To simplify the discussion we will assume that
\eqn\kjju{l=0\ ({\rm mod}\  4)\ .}
For any   solution\ $\varphi(r,\theta)$\
of the Sine-Gordon
equation in the region \ \gfbn\  we can construct the following matrix
\eqn\hsgdf{{\bf T}^{\theta}(\alpha)=
\lim_{\scriptstyle a\to 0\atop\scriptstyle R\to+\infty}
e^{\frac {i m R }{2} \sinh \alpha
\sigma_3}{\bf U}\overleftarrow{P}
\exp[\int_{a}^{R}d r {\cal A}_{r}(r,\theta,\alpha)\big]\ .}
Here
$${\bf U}=\frac{1}{\sqrt2}\left[\matrix{1&-i\cr -i&1}\right] .$$
If the functions\ $\pi(r),\varphi(r)$\
tend to their limiting values\ \hdg\  fast enough,
the standard arguments\ \col ,\fa\  allow one to prove that the
matrix\ ${\bf T}^{\theta}(\alpha)$\ exists and satisfies the
conditions:

{\bf 1.}  The  dynamical equation
\eqn\hdfg{{\bf T}^{\theta}(\alpha)={\bf T}(\alpha-\theta)\ .}

{\bf 2.} The matrix \ ${\bf T}(\alpha)$\  has the following analytical
structure
\eqn\gsx{{\bf T}(\alpha)=
\left[\matrix{i\  T_+(\alpha)&i\   T_-(\alpha)
\cr T_+(\alpha+i \pi)&-T_-(\alpha+i\pi)}\right]\ .}
Here  \ $T_a(\alpha)\  (a=\pm$)\ are analytical functions in the
strip\ $0\leq \Im m\  \alpha\leq \pi$. Hereinafter we shall call
them  the Jost functions.

{\bf 3.} For real $\alpha$\ the Jost functions satisfy the conjugation
condition
\eqn\hdf{i\  T^*_{a}(\alpha)=\C^{ab} T_b(\alpha+i\pi)\ .}
Here we use the notation\ $\C^{ab}=\C_{ab}=\delta_{a+b}\ .$

{\bf 4.} The determinant of the matrix\ ${\bf T}(\alpha)$ is equal to one:
\eqn\hsgdfa{\C^{ab}T_a(\alpha+i\pi)T_b(\alpha)=i\ .}

\noindent
We  restrict the class of initial
data by the following additional
requirements:

{\bf 5.} The function \ $T_+(\alpha)$\ does not have zeros in the strip
of analyticity\ $0\leq \Im m\ \alpha\leq \pi$.

{\bf 6.} If \ $Re\ \alpha\to \pm \infty\ ,\
0\leq\Im m\ \alpha\leq\pi$ \ then
\eqn\jdhdo{[T_+(\alpha)]^{-1}=
o(\alpha^{-\frac{1}{2}}e^{\frac{|Re\alpha|}{2}})\ .}
\noindent
It is necessary to point out that the analytical
condition {\bf 2} can be considered as a
constraint on asymptotic behavior at \ $r\sim R\to +\infty$
of the initial functions\ $\pi(r),\varphi(r)$\ \hdg\  .
At the same time the
requirement\ {\bf 6}  determines their  permissible behavior in
other asymptotic area \ $r\sim a\to 0$\ .  Let us also note that
the additional requirements {\bf 5,6}  considered as constraints on
the initial data are compatible with the
dynamical equation \ \hdfg\ .

Using the properties ({\bf 1-6}) of
the Jost functions it is easy to prove the dispersion
relation:
\eqn\gsfds{\frac{T_-(\alpha)}{T_+(\alpha)}=
\int_{-\infty}^{+\infty}\frac{d \delta}{2 \pi}\
\frac{[T_+(\delta) T_+(\delta+i \pi)]^{-1}}
{\sinh(\delta-\alpha-i 0)}.}
Introduce the function
\eqn\ksjh{\lambda(\alpha)=-i\  \ln T_+(\alpha)\ .}
{}From the above, the formula
\hsgdf\  determines  the transformation
\eqn\gsd{Scat:\ \varphi(r,\theta)
\to\lambda^{\theta}(\alpha)=\lambda(\alpha-\theta).}
It generalizes the transformation\ \jhggg\  for the Klein-Gordon model
and in the weak-coupling limit\ $b\to 0$\
\eqn\fsdd{\lambda^{SG}(\alpha)\to\frac{b}{4}
\lambda^{KG}(\alpha)+const.}
Note that the solution of the Sine-Gordon equation
in the region
$x>|t|>0$\  is reduced to applying
the direct and inverse transformation \ $Scat$\ \gsd\ .

Let us  describe now the canonical Poisson
structure\ \hdy\  in terms
of the Jost functions.
Using the familiar method of calculation\ \fa ,
one can show that the functions \ $T_{a}(\alpha)$\ form
a quadratic Poisson algebra:
\eqn\lso{\{T_a(\alpha_1),
T_b(\alpha_2)\}=r_{ab}^{cd}(\alpha_1-\alpha_2)
T_d(\alpha_2)T_c(\alpha_1),
\ \  -\pi< \Im m\  (\alpha_1-\alpha_2)<\pi\ .}
The  matrix\ $r_{ab}^{cd}(\alpha)$\ has the following nontrivial
elements:
\eqn\hgdf{\eqalign{&r_{++}^{++}
=r_{--}^{--}=\frac{b^2}{16}\tanh \frac
{\alpha}{2}\ ,\cr
&r_{+-}^{+-}=r_{-+}^{-+}=-\frac{b^2}{16}\  {\rm P.V.}\coth \frac
{\alpha}{2}\ ,\cr
&r_{+-}^{-+}=r_{-+}^{+-}=\frac{b^2}{16}\ {\rm P.V.}
\frac{2}{\sinh\alpha}\  .}}
These formulas are equivalent to the  simple Poisson
bracket for the  function\ $\lambda(\alpha)$\ \ksjh
\eqn\rwewq{\{\lambda(\alpha_1),\lambda(\alpha_2)\}
=-\frac{b^2}{16}
\tanh\frac{\alpha_1-\alpha_2}{2},
\ \ \ -\pi< \Im m\ (\alpha_1-\alpha_2)<\pi\ .}
Hence we have exactly the same Poisson
structure as for the Klein-Gordon equation.

The quantization technique for the Poisson
algebra\ \lso\  has been
developed in \ \luk .
In principle, it is a natural generalization
of ideas considered in Sec.\  2.
Here we only  note  that  the quantum operators\ $T_a(\alpha)$\
satisfy the condition\ \hsgdfa\  and the  commutation relation:
\eqn\gsfd{T_a(\alpha_1)T_b(\alpha_2)=
R_{ab}^{cd}(\alpha_1-\alpha_2)
T_d(\alpha_2) T_c(\alpha_1).}
The matrix\ $R_{ab}^{cd}$\ reads explicitly
\eqn\fsdq{\eqalign{&R_{++}^{++}=R_{--}^{--}=R(\alpha),\cr
&R_{+-}^{+-}=R^{-+}_{-+} =-R(\alpha)\frac{\sinh\alpha\nu}
{\sinh(i \pi+\alpha)\nu},\cr
&R_{+-}^{-+}=R^{+-}_{-+}=R(\alpha)\frac{\sinh i\pi\nu}
{\sinh(i \pi+\alpha)\nu}.}}
Here the function\ $R(\alpha)$\ is represented by :
\eqn\bvvcre{R(\alpha)=\frac{\Gamma(\nu)
\Gamma(1-\frac{i\alpha\nu}{\pi})}
{\Gamma(\nu-\frac{i\alpha\nu}{\pi})}
\prod_{p=1}^\infty\frac{R'_p(-\alpha)
R'_p(i\pi+\alpha)}{R'_p(0)R'_p(i\pi)},}
$$R'_p(\alpha)=\frac{\Gamma(2 p\nu+\frac{i\alpha\nu}{\pi})
\Gamma(1+2 p\nu+\frac{i\alpha\nu}{\pi})}
{\Gamma((2 p+1)\nu+\frac{i\alpha\nu}{\pi})
\Gamma(1+(2 p-1)\nu+\frac{i\alpha\nu}{\pi})},$$
and the parameter\ $\nu$\
is connected with the constant of interaction\ $b$\
in the following way:
\eqn\mak{\nu=1-\frac{b^2 \hbar}{8\pi} .}
One can point out  that quantization of the
dispersion relation\ \gsfds\
leads to
the free field representation of the algebra\ \gsfd\ .

As already  discussed in the Sec.\ 2 the vacuum correlation functions
$$<vac|T^{\theta_n}_{a_n}(\alpha_n)...
T^{\theta_1}_{a_1}(\alpha_1)|vac>\equiv
{\cal F}_{a_1...a_n}(\alpha_1-\theta_1,
...\alpha_n-\theta_n)$$
can be expressed in terms of traces over
the space of angular quantization\
$\pi_Z$.
Using the notation of \ \luk\  one
can write them in the following form:
\eqn\gsfddar{{\cal F}_{a_1...a_n}(\alpha_1,...\alpha_n)
=
tr_{\pi_Z}\big[e^{2\pi i K}Z'_{a_n}(\alpha_n)...
Z'_{a_1}(\alpha_1)\big]\ .}
In \ \luk\  the integral
representation for these function has
been constructed. Here we present  explicitly only
the simplest nontrivial one.
\eqn\hfgdf{F_{ab}(\alpha_1,\alpha_2)=
\C_{ab}\frac{e^{\frac{a\nu}{2}(\alpha_1-
\alpha_2-i\pi)}}{4\nu\cos\pi\nu}
\frac{\sinh\nu(\alpha_1-\alpha_2+i\pi)}
{\cosh\frac{ \alpha_1-\alpha_2}{2}}
\frac{G'(\alpha_1-\alpha_2)}{G'(-i\pi)} \ ,}
where
\ $G'(\alpha)$\ is given by\
\eqn\jfhhgp{G'(\alpha)=
\exp\big[\int_{0}^{+\infty}\frac{d t} {t}
\frac {\sinh^2t (1-i \frac{\alpha}
{\pi}) }{\sinh 2 t \cosh t}
\frac{\sinh\frac{t(1-\nu)}{\nu}}{\sinh\frac{t}{\nu}}\big]\ .}

So far we have considered only  region\ $x>|t|>0$.
Let us discuss now another space-like
region in Minkowski space
\eqn\jsgd{x<-|t|<0\  .}
The coordinates in \jsgd\  can be chosen as\ \jshg ,
where \ $ \theta=-\pi i + \zeta,\ -\infty<\zeta<\infty$.
One can introduce the Jost functions  \
$ Q^a_{\zeta}(\alpha)$\
in an analogous way as
for the region\ $x>|t|>0$:
\eqn\kijh{{\bf Q}_{\zeta}(\alpha)={\bf Q}(\alpha-\zeta)=
\lim_{\scriptstyle a\to 0\atop\scriptstyle R\to+\infty}
\overleftarrow{P}
\exp[\int_{R}^{a}d r {\cal A}_{r}(r,-i\pi+\zeta,\alpha)\big]
{\bf U}^{-1}e^{\frac {i  m R }{2}
\sinh \alpha
\sigma_3},}
and
\eqn\hsgdf{{\bf Q}(\alpha)=
\left[\matrix{ Q^+(\alpha)&i
Q^+(\alpha+i\pi)
\cr \  Q^-(\alpha)&-i\ Q ^-(\alpha+i\pi)}\right]\ .}
Then  standard  arguments\
\hdgf\
show
that  vacuum
correlation functions of the operators\
$T^{\theta}_a(\alpha)$\ and\ $ Q_{\zeta}^b(\alpha)$\
can be expressed in terms of  \gsfddar:
\eqn\bcvs{\eqalign{
&<vac|
T^{\theta_n}_{a_n}(\alpha_{n})...T^{\theta_1}_{a_1}(\alpha_{1})
Q_{\zeta_1}^{b_1}(\alpha_{m+1})...
Q_{\zeta_{m}}^{b_m}(\alpha_{n+m })|vac>=
\C^{b_1c_1}...\C^{b_mc_m}\times\cr&\times{\cal F}
_{a_1....a_nc_1...c_m}(\alpha_1-\theta_1,...,
\alpha_n-\theta_n,
i\pi+\alpha_{n+1}-\zeta_1,...,i\pi+\alpha_{n+m}-
\zeta_{m})\ .}}
Particularly,  the functions
$${\cal F}_{a_1...a_nc_{1}...c_{n}}(\alpha_1,...
\alpha_n,i\pi+\alpha_n,
...,i\pi+\alpha_1)$$
describe the vacuum correlators of the  elements of the monodromy
matrix
\eqn\mcnv{{\bf M}(\alpha)={\bf T}(\alpha) {\bf Q}(\alpha) }
in the Sine-Gordon model.
The technique developed in the work \ \luk\
also provides  calculations of
matrix elements of arbitrary combination of
the Jost functions for
soliton asymptotic states.
In the
notation of \ \luk\  they are  expressed in terms of
functions:
\eqn\hxvc{
tr_{\pi_Z}\big[e^{2\pi i K}Z'_{a_n}(\alpha_n)...
Z'_{a_1}(\alpha_1)
Z_{b_1}(\beta_1)...Z_{b_k}(\beta_k)\big],}
which admit integral representations as well as\ \gsfddar .

\newsec{Conclusion}

The quantum direct scattering problem for the
Sine-Gordon model admits a complete solution.
Now we should  try to solve the inverse scattering problem,
which is equivalent to determining   correlation functions
of local operators. At present there is only an indirect
way to do this:
The asymptotic of functions\
 \hxvc\  at \ $Re\  \alpha_k\to\pm\infty$ \
defines the form-factors of all local operators
in the theory\ \luk.
Knowing the form-factors one
can reconstruct the correlation functions
in the familiar manner\
\ref\gsfd{B.Berg, M. Karowski and P.Weisz,
Phys. Rev. D19 (1979) 2477},
\ref\sm{F.A. Smirnov, Form-Factors
in Completely Integrable Models of
Quantum Field Theory, World Scientific (1992)},
\ref\mus{G. Mussardo, Phys. Reports 218 (1992) 215}.
Hopefully, there is a more effective
and direct way to derive physical  correlation functions
in terms of only  correlators of the Jost functions\ \bcvs .
{}From this point of view,
it will be very useful to analyze the free fermion point
$\ b^2\hbar=4\pi$\
\ref\lec{A. LeClair, Spectrum Generating
Affine Algebras and Correlation Functions in
Massive Field Theory,
Cornell preprint  CLNS 93/1220 (1993) (hep-th/9305110)}.
\centerline{}

\centerline{\bf Acknowledgments}

I am grateful to  A.B. Zamolodchikov for
extremely useful discussions. I am also pleased to thank
A. LeClair  for discussions and hospitality
at Cornell University where the work
was finished.
This work was supported by grant DE-FG05-90ER40559.

\listrefs

\end